\documentclass[a4paper,11pt]{article}
\usepackage{pos}

\usepackage[utf8]{inputenc}
\usepackage{amsmath}

\usepackage{amssymb}
\usepackage{graphicx}% Include figure files
\usepackage[capitalize]{cleveref}
\graphicspath{{./graphics/}}
\newcommand{\be}{\begin{equation}}
	\newcommand{\ee}{\end{equation}}

\newcommand{\beq}{\begin{equation}}
	\newcommand{\eeq}{\end{equation}}
\newcommand{\bea}{\begin{eqnarray}}
\newcommand{\eea}{\end{eqnarray}}

\usepackage[T1]{fontenc}
\usepackage{tikz-cd}
\usepackage[british]{babel}
\usepackage{layout}
\usepackage{amsmath}
\usepackage{amstext, amsthm, simplewick, amsfonts}
\usepackage{framed}

\title{Gravitational Form Factors and the QCD Dilaton at Large Momentum Transfer}

\ShortTitle{GFF and the Perturbative Dilaton}

\author{{Claudio Corian\`o}$^{1,2}$\footnote{Presenter}\,  Stefano Lionetti $^{1}$, Dario Melle $^{1,2}$, Riccardo Tommasi$^{1}$, Leonardo Torcellini$^{1}$\\
$^1$ Dipartimento di Matematica e Fisica,
Universit\`a del Salento\\
and INFN Sezione di Lecce, 73100, Lecce, Italy\\
$^2$ National Center for HPC, Big Data and Quantum Computing}

\abstract{We investigate the hard scatterings of hadronic matrix elements corresponding to hadronic gravitational form factors (GFFs) of the pion and proton using QCD factorization, applying conformal field theory (CFT) tools. These GFFs are key to understanding quark and gluon angular momentum via their connection to DVCS moments. The core object is the non-Abelian \( TJJ \) 3-point function, which shows an anomaly-induced dilaton exchange in the \( t \)-channel. We analyze quark, ghost, and gauge-fixing effects through a CFT-based decomposition and propose a parameterization useful for future DVCS studies at the Electron-Ion Collider. The dilaton interaction is interpolated by a conformal anomaly form factor, defined in the nonconformal case, which is constrained by a (dilaton) sum rule. }

\FullConference{Corfu Summer Institute 2024 "School and Workshops on Elementary Particle Physics and Gravity"\\
		(CORFU2024)\\
		31 August - 28 September, 2024\\
		Corfu, Greece}

\begin{document}

\maketitle
\section{Introduction}
\label{gen1}

The experimental study of gravitational form factors (GFFs) of the proton and pion provides nonperturbative insights into their coupling to the QCD energy-momentum tensor (EMT), revealing information on the distribution of energy, spin, pressure, and shear forces (see \cite{Polyakov:2018zvc,Teryaev:2016edw,Burkert:2023wzr}).\\
The interest in the GFF of the proton \cite{Tong:2022zax,Tanaka:2022wrr,Hatta:2018sqd,} is part of  efforts undertaken in the analysis of such hadronic matrix elements \cite{Tong:2022zax}, connected with the measurements of the Deeply Virtual Compton Scattering (DVCS) cross section and of generalized parton distributions \cite{Ji:1998pc,Radyushkin:1997ki} at JLab and at the 
Electron-ion collider (EIC) at Brookhaven.
Our work, in this case, is concerned about the role of the conformal anomaly and of its perturbative manifestation 
in QCD. Analysis of anomalies at parton level, particularly in the chiral case, have been reconsidered in several works 
\cite{Tarasov:2025mvn,Tarasov:2021yll,Bhattacharya:2023wvy,Bhattacharya:2022xxw,Castelli:2024eza}.  
GFFs are defined through matrix elements of the EMT between hadron states. These expansions reveal how the proton’s internal structure relates to its energy, momentum, and stress distributions. For a spin-$\frac{1}{2}$ hadron, the matrix elements of the EMT take the form
\begin{align}
\langle p^\prime,s^\prime| T_{\mu\nu}(0) |p,s\rangle = \bar u{ }^\prime\biggl[
A(t)\,\frac{\gamma_{\{\mu} P_{\nu\}}}{2}
+ B(t)\,\frac{i\,P_{\{\mu}\sigma_{\nu\}\rho}\Delta^\rho}{4 M}
+ D(t)\,\frac{\Delta_\mu\Delta_\nu-g_{\mu\nu}\Delta^2}{4M}
+ M\,\sum_{\hat{a}} {\bar c}^{\hat{a}}(t)\,g_{\mu\nu} \biggr]u,
\label{fund1}
\end{align}
where $u(p)$, $\bar{u}(p')$ are the proton spinors, $P = (p + p')/2$, $\Delta = p' - p$, $t = \Delta^2$, and $M$ the proton mass. The notation $\gamma^{\{\mu} P^{\nu\}}$ denotes symmetrization over indices.\\
Using the Gordon identity, the EMT decompositions for quarks ($\hat{a}=q$) and gluons ($\hat{a}=g$) are
\bea
\langle p^\prime,s^\prime| T_{\mu\nu}^{\hat{a}}(0) |p,s\rangle = \bar u^\prime\biggl[
A^{\hat{a}}(t)\,\frac{P_\mu P_\nu}{M}
+ J^{\hat{a}}(t)\,\frac{i\,P_{\{\mu}\sigma_{\nu\}\rho}\Delta^\rho}{2 M}
+ D^{\hat{a}}(t)\,\frac{\Delta_\mu\Delta_\nu-g_{\mu\nu}\Delta^2}{4 M}
+ M\,\bar{c}^{\hat{a}}(t)g_{\mu\nu} \biggr]u,
\label{Eq:EMT-FFs-spin-12-alternative}
\eea
where the form factors satisfy $A^{\hat{a}}(t) + B^{\hat{a}}(t) = 2 J^{\hat{a}}(t)$.\\
For a spin-0 hadron, the matrix element is
\bea
\label{fund2}
\langle p^{\,\prime}| \hat{T}_{\mu\nu}(0) |p\rangle =
\biggl[2P_\mu P_\nu\, A(t)
+ \frac{1}{2}(\Delta_\mu\Delta_\nu - g_{\mu\nu}\Delta^2)\, D(t)
+ 2m^2\,\bar{c}(t)\,g_{\mu\nu}\biggr].
\eea

The EMT can be accessed through generalized parton distribution (GPD) studies \cite{Ji:1996ek,Radyushkin:1996nd,Radyushkin:1996ru,Ji:1996nm,Collins:1996fb,Radyushkin:1997ki,Vanderhaeghen:1998uc} in hard exclusive processes, providing information on the mass and spin structure of hadrons \cite{Kobzarev:1962wt,Pagels:1966zza,Ji:1998pc,Radyushkin:2000uy,Goeke:2001tz,Diehl:2003ny,Belitsky:2005qn}. 
Along with the $D$-term \cite{Polyakov:1999gs}, these form factors offer a detailed tomography of the proton. Specifically, $A(t)$ encodes the momentum distribution, with $A(0)=1$ corresponding to the total proton momentum.
$B(t)$ describes the angular momentum distribution. It satisfies $B(0) = 0$, reflecting the vanishing anomalous gravitomagnetic moment, consistent with the Poincaré group constraints.
In contrast, $D(t)$ at $t=0$ is unconstrained and characterizes internal forces and stress distributions. \\
The combination $A(t)+B(t)$ at $t=0$ yields the total angular momentum through Ji’s sum rule \cite{Ji:1996ek}
\beq
J_q + J_g = \frac{1}{2} [A_q(0) + B_q(0)] + \frac{1}{2} [A_g(0) + B_g(0)] = \frac{1}{2}.
\eeq
In this talk our concern will be to establish a link with perturbative QCD analysis of the factorization formula for this matrix element at large momentum transfers, similarly to the analysis presented in \cite{Tong:2022zax}. We will be focusing, at the moment, only on the discussion of the structure of its hard scattering contribution around the conformal limit. Details of this analysis can be found in \cite{Coriano:2024qbr}. An in-depth discussion of the sum rules for anomaly form factors, which characterize the dilaton interactions in QCD at large momentum transfers, is provided in two recent works \cite{Coriano:2025fom,Coriano:2025ceu}, to which we refer the reader for further details. Extensions of this analysis, focusing on the conformal properties of hard scattering processes—and their modifications around the conformal limit—in hadronic matrix elements are currently underway.

\section{ Their relation to DVCS }

Gravitational form factors (GFFs) are related to DVCS of an electron off a nucleon ($eN\to e^\prime N^\prime \gamma$) with a final-state photon. DVCS involves a high-energy electron scattering off a hadron (such as a proton or pion) via virtual photon exchange, followed by real photon emission. Analyses with other neutral currents are also possible \cite{Amore:2004ng}.\\
The process interpolates between the soft region, where QCD sum rules apply \cite{Coriano:1993mr} and the Feynman mechanism dominates, and the inelastic high-energy region \cite{Coriano:1998ge,Sterman:1997sx}.\\
Access to EMT form factors is achieved through generalized parton distributions (GPDs) 
\cite{Ji:1996ek,Radyushkin:1996nd,Radyushkin:1996ru,Ji:1996nm,Collins:1996fb,Radyushkin:1997ki,Vanderhaeghen:1998uc}, which describe hard-exclusive reactions such as DVCS, or exclusive meson production $eN\to e^\prime N^\prime M$.\\
For the nucleon, the second Mellin moments of unpolarized GPDs yield the EMT form factors $A$, $B$, and $D$:
\begin{align}
\int_{-1}^1{\rm d}x\;x\, H^a(x,\xi,t) &= A^a(t) + \xi^2 D^a(t) \,, \quad
\int_{-1}^1{\rm d}x\;x\, E^a(x,\xi,t) = B^a(t) - \xi^2 D^a(t)\,.
\label{Eq:GPD-Mellin}
\end{align}
Here, $H$ and $E$ parameterize light-cone amplitudes with nonforward kinematics, describing the removal and reinsertion of a parton with momentum fractions $(x-\xi)$ and $(x+\xi)$, respectively. $\Delta$ ($\Delta^2=t$) denotes the momentum transfer, while $\xi$ is a second scaling variable. Through their Mellin moments, GPDs encode information about the GFFs.\\
The importance of GFFs for understanding hadron structure has driven experimental efforts. Initial determinations of proton quark~\cite{Burkert:2018bqq} and gluon~\cite{Duran:2022xag} GFFs have been achieved via DVCS and $J/\psi$ photoproduction measurements. 
Progress on the pion GFFs has been achieved using data from the Belle experiment at KEKB \cite{Belle:2015oin,Savinov:2013hda,Kumano:2017lhr}.\\
Further advances are anticipated from upcoming experiments at the JLab 12 GeV program~\cite{JeffersonLabHallA:2022pnx,CLAS:2022syx} and the future Electron-Ion Collider (EIC)~\cite{AbdulKhalek:2021gbh}.

\section{Factorization at large momentum transfer}
At sufficiently large momentum transfer, the gravitational form factor (GFF) can be described within a factorization framework (see Figs. 1, 2), where the $TJJ$ vertex (Fig. 3) —central to our analysis—plays a key role. In this context, one can employ conventional collinear factorization using distribution amplitudes or adopt a modified scheme that incorporates Sudakov effects, as discussed in the past in studies of the proton’s electromagnetic form factor~\cite{Li:1992nu}.\\
\noindent
The light-front Fock state expansion offers a natural and powerful representation of hadronic structure~\cite{Brodsky:1997de}. This expansion formally comprises an infinite series of Fock states, each associated with specific partonic configurations and corresponding light-front wave functions. However, for exclusive processes involving large momentum transfers, general power-counting arguments indicate that the leading contributions stem from the lowest Fock states—those with the minimal number of partons and the least orbital angular momentum~\cite{Matveev:1973ra, Brodsky:1973kr, Ji:2003fw}.\\
\noindent 
In the factorization approach the dominant contribution arises from the leading three-valence-quark Fock state of the proton, following the procedure outlined in~\cite{Ji:2002xn, Ji:2003yj, Tong:2022zax}.
Generally, hadronic and partonic matrix elements are connected through hadronic wave functions integrated over the partons’ fractional momenta in the hard scattering amplitude. For example, in the case of the electromagnetic form factor of the pion with initial momentum \( p \) and final momentum \( p' \), one has~\cite{Sterman:1997sx}

\begin{equation}
(p' + p)_\mu F_\pi(q^2) = \langle \pi(p') | J_\mu(0) | \pi(p) \rangle,
\end{equation}
\noindent which, at large \( q^2 \), factorizes as

\begin{equation}
F_\pi(q^2) = \int_0^1 dx \, dy \, \phi_\pi(y, \mu^2)\, T_H(y, x, q^2, \mu^2)\, \phi_\pi(x, \mu^2).
\end{equation}
\noindent
Here, \( J^\mu \) is the electromagnetic current for the valence quarks in the pion, and \( \phi_\pi(x, \mu^2) \) represents the distribution amplitude, describing the momentum fraction \( x \) carried by the quark (and \( 1 - x \) by the antiquark) in the lowest Fock state. The factorization scale \( \mu \) evolves under the renormalization group via the Efremov-Radyushkin-Brodsky-Lepage (ERBL) equation~\cite{Lepage:1980fj, Efremov:1979qk}.\\
A similar structure applies to GFFs. However, the identification of the conformal anomaly contribution within the hard scattering kernel \( T_H \) is substantially more intricate.
\subsection{Extracting the perturbative $TJJ$ and the dilaton pole}
 The corresponding process for a GFF is illustrated in Fig. 1 for the pion and in Fig. 2  for the proton. The amplitudes are mediated by the $TJJ$ vertex, appearing at $O(\alpha_s^2)$ and $O(\alpha_s^3)$ respectively, in the hard subprocess. In our approach we investigate the full tensor structure of the interaction, proposing a decomposition 
 valid generically, and identifying a conformal anomaly form factor which exhibits a sum rule in its trace sector only if the vertex is kept uncontracted. Details are given in two recent works covering chiral, conformal and gravitational anomaly-mediated interactions \cite{Coriano:2025fom,Coriano:2025ceu} that exhibit similar features.\\
 In the case of the perturbative $TJJ$ correlator in the conformal limit, the dilaton interaction manifests through the exchange of a pole with a nonvanishing residue only under light-cone kinematics, requiring on-shell gluons. More generally, a dilaton interaction is characterized by a dispersive cut, constrained by a sum rule. Thus, an asymptotic dilaton state, appearing as a pole, saturates the spectral density only in the presence of light-cone dominance of the perturbative interaction. The sum rule, a hallmark of these anomaly-mediated interactions, can be shown to be preserved at one loop across all kinematical configurations (i.e., both with virtual or on-shell gluons in the $TJJ$ and with massive quarks).\\
  In the conformal limit (massless fermions) and for on-shell gluons, as mentioned, the spectral density of the anomaly form factor becomes localized, manifesting as a simple pole in the $t$-channel. Consequently, the description of these interactions naturally leads to a nonlocal action. Such nonlocal actions have been extensively studied over several decades, particularly in the context of gravity, where correlators involving multiple insertions of the stress-energy tensor coupled to external classical gravitational fields have been analyzed for quantum conformal matter \cite{Coriano:2020ees}. In the QCD case, the $TJJ$ analysis is modified by the presence of a gauge-fixing sector, which breaks the conformal symmetry of QCD already at tree level. 
 
  \subsection{The tensorial decomposition of the interaction}
We propose a tensorial decomposition of the interaction, which efficiently encodes the conformal constraints at one-loop. The formalism has been developed in the analysis of such constraints in momentum space \cite{Bzowski:2013sza}, taking the form of conformal Ward identities (CWIs), accompanied by conservation and anomalous trace Ward identities. In the perturbative case, the abstract CFT formulation of this correlator \cite{Bzowski:2013sza} needs to be modified by the inclusion of Slavnov-Taylor identities which are less restrictive compared to the ordinary Ward identities imposed in the Abelian case, due to gauge fixing \cite{Coriano:2024qbr}. \\
\noindent
 The anomaly contribution arises at NLO in the factorizaton formula as indicated in Fig. 1 (right), where the two gluons in the $TJJ$ vertex couple to the hard scattering amplitude in all possible ways. In the pion case, for instance, non-anomalous contributions are also present, appearing at leading order in \( \alpha_s \) (Fig. 1 (left)), by a direct insertion of the EMT, 
 and extend to \( \mathcal{O}(\alpha_s^2) \), with the inclusion of box-type contributions. In the proton case the anomaly contribution appears at  \( \mathcal{O}(\alpha_s^3) \). The additional suppression, in this case, is due to the presence of three collinear quarks in the factorization picture, which requires the hard rescatter of two gluons at leading order. 

\begin{figure}[t]
    \centering
    \includegraphics[width=0.4\textwidth]{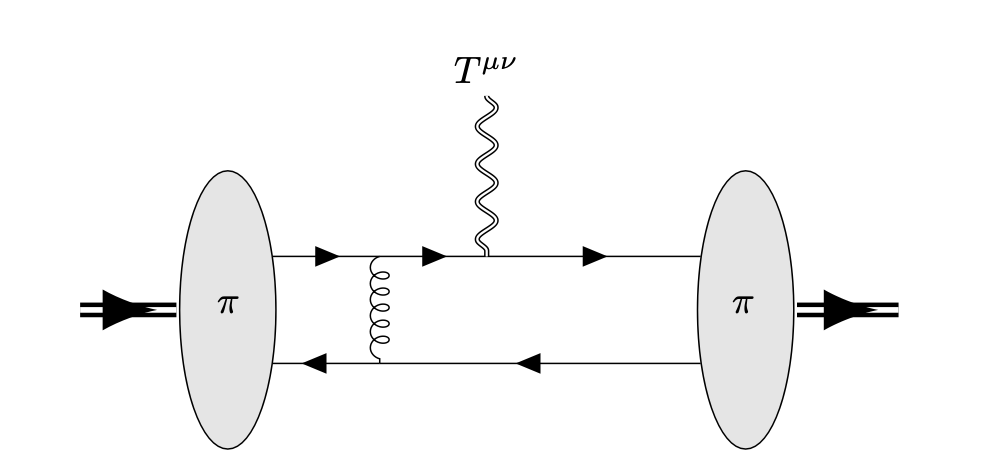}
    \includegraphics[width=0.4\textwidth]{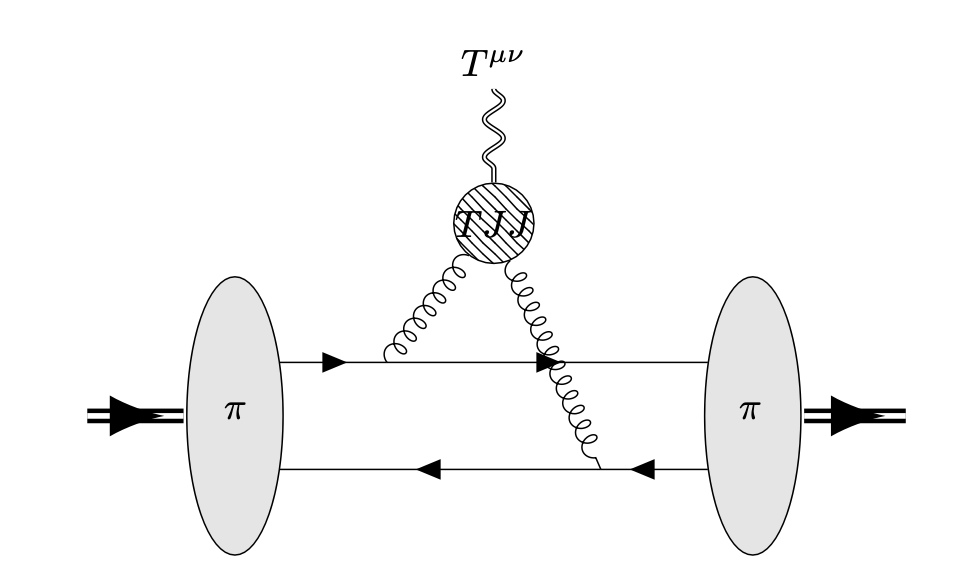}
       \caption{Typical leading (left) and NLO contributions (right) to the GFF of the pion. The anomaly mediated interaction, in this case, appears at $O(\alpha_s^2)$. }
    \label{fig:your_label}
\end{figure}
\noindent 

\section{The hadronic GFF  and the conformal anomaly}
The non-Abelian \( TJJ \) correlator is expanded off-shell using sector decomposition into longitudinal, transverse-traceless (tt), and trace components. This decomposition has been used quite generally in $CFT_p$ \cite{Bzowski:2013sza} and makes it possible to isolate anomaly form factors in a consistent way.\\
We can move away form the conformal limit and still use the same decompostion, by including a nonzero fermion mass in the correlator and re-analizing the decomposition.
In this decomposition, as described in \cite{Coriano:2024qbr}, while the quark sector is fixed by CWIs, the gluon contributions are added perturbatively. \\
Typical contributions are shown in Fig. 3. 
The final parameterization captures both quark and gluon sectors and isolates the anomaly pole, interpreted as due to a dilaton exchanged in the $t$-channel. 
Each operator is decomposed as
\beq
T^{\mu\nu}(p) = t^{\mu\nu}(p) + t_{\text{loc}}^{\mu\nu}(p), \quad
J^{a\mu}(p) = j^{a\mu}(p) + j^{a\mu}_{\text{loc}}(p),
\eeq
where transverse (or transverse traceless) and longitudinal parts are defined using projectors, and $a, b$ denote colour indices.  
	\begin{align}
		\pi^{\mu}_{\alpha} & = \delta^{\mu}_{\alpha} - \frac{p^{\mu} p_{\alpha}}{p^2}, \\
		%%%%%%%%%%%%%%%%%%%%%%%%%%%%%%%
		\Pi^{\mu \nu}_{\alpha \beta} & = \frac{1}{2} \left( \pi^{\mu}_{\alpha} \pi^{\nu}_{\beta} + \pi^{\mu}_{\beta} \pi^{\nu}_{\alpha} \right) - \frac{1}{d - 1} \pi^{\mu \nu}\pi_{\alpha \beta}\label{TTproj}, \\
\end{align}
$t^{\mu\nu}(p)$, for example, is a transverse-traceless operator in momentum space. The {\em local} contribution 
 $t_{\text{loc}}^{\mu\nu}(p)$ is defined as a the remainder of the decomposition.  Similarly,  
$j^{a\mu}(p)$ is the transverse part of the gauge current with respect to a certain momentum $p$
\begin{align}
		\label{loct}
		t^{\mu_i\nu_i}(p_i)&=\Pi^{\mu_i\nu_i}_{\alpha_i\beta_i}(p_i)\,T^{\alpha_i \beta_i}(p_i), &&t_{loc}^{\mu_i\nu_i}(p_i)=\Sigma^{\mu_i\nu_i}_{\alpha_i\beta_i}(p)\,T^{\alpha_i \beta_i}(p_i),\\
		j^{a_i \, \mu_i}(p_i)&=\pi^{\mu_i}_{\alpha_i}(p_i)\,J^{a_i \, \alpha_i }(p_i), &&\hspace{1ex}j_{loc}^{a_i \, \mu_i}(p_i)=\frac{p_i^{\mu_i}\,p_{i\,\alpha_i}}{p_i^2}\,J^{a_i \, \alpha_i}(p_i).
	\end{align}

The projectors are designed to bring the CWIs to a manageable form, with the inclusion of a minimal number of form factors, introduced in the transverse traceless sector. 
The full 3-point function is decomposed into eight terms
\begin{align}
\langle T^{\mu\nu} J^{a\alpha} J^{b\beta} \rangle &= 
\langle t^{\mu\nu} j^{a\alpha} j^{b\beta} \rangle 
+ \langle T^{\mu\nu} J^{a\alpha} j^{b\beta}_{\text{loc}} \rangle 
+ \langle T^{\mu\nu} j^{a\alpha}_{\text{loc}} J^{b\beta} \rangle 
+ \langle t^{\mu\nu}_{\text{loc}} J^{a\alpha} J^{b\beta} \rangle \nonumber \\
&\quad 
- \langle T^{\mu\nu} j^{a\alpha}_{\text{loc}} j^{b\beta}_{\text{loc}} \rangle 
- \langle t^{\mu\nu}_{\text{loc}} j^{a\alpha}_{\text{loc}} J^{b\beta} \rangle 
- \langle t^{\mu\nu}_{\text{loc}} J^{a\alpha} j^{b\beta}_{\text{loc}} \rangle 
+ \langle t^{\mu\nu}_{\text{loc}} j^{a\alpha}_{\text{loc}} j^{b\beta}_{\text{loc}} \rangle .
\end{align}
This approach systematically exposes the anomalous parts of the off-shell gravitational form factors after renormalization. It can be performed either generally, without reference to free field theory realizations of the CFT, as discussed in \cite{Bzowski:2018fql}, using special integrals of Bessel functions, or by applying ordinary dimensional regularization to Feynman master integrals if we resort to free field theory. In a CFT, when dealing with correlators of vector currents and stress-energy tensors, the two approaches can be exactly matched. In this second case, the form factors of the transverse traceless sector
\beq
\langle t^{\mu\nu} j^{a\alpha} j^{b\beta} \rangle , 
\eeq
which are the building blocks of the decomposition, are expressed in terms of ordinary Feynman integrals. 
The standard \( CFT_p \) approach, however, cannot be applied directly to a gauge-fixed theory such as QCD, due to the need to replace the ordinary Ward identities satisfied by the gluon currents with Slavnov-Taylor identities, which are less restrictive. As a result, conformal symmetry is broken, and contributions involving virtual gluons must be treated perturbatively. Mass corrections in the quark sector are also organized according to the same scheme, leading to results that differ significantly from those discussed in \( CFT_p \) \cite{Bzowski:2018fql}. We have derived the decomposition
\begin{equation}
\begin{aligned}
\langle T^{\mu \nu}(q) J^{a\alpha}(p_1) J^{b\beta}(p_2) \rangle = &\ \langle t^{\mu \nu}(q) j^{a\alpha}(p_1) j^{b\beta}(p_2) \rangle + \langle t^{\mu \nu}(q) j_{\text{loc}}^{a\alpha}(p_1) j^{b\beta}(p_2) \rangle_g + \langle t^{\mu \nu}(q) j^{a\alpha}(p_1) j_{\text{loc}}^{b\beta}(p_2) \rangle_g  \\
&+ 2 \mathcal{I}^{\mu \nu \rho}(q) \left[ \delta_{[\rho}^{\beta} p_{2 \sigma]} \langle J^{a\alpha}(p_1) J^{b\sigma}(-p_1) \rangle + \delta_{[\rho}^{\alpha} p_{1 \sigma]} \langle J^{b\beta}(p_2) J^{a\sigma}(-p_2) \rangle \right]  \\
&+ \frac{1}{3 q^2} \hat{\pi}^{\mu \nu}(q) \left[ \mathcal{A}^{\alpha \beta ab} + \mathcal{B}^{\alpha \beta ab}_g \right],
\end{aligned}
\end{equation}
where the traceless sector is given by
\begin{equation}
\begin{aligned}
\langle T^{\mu \nu}(q) J^{a\alpha}(p_1) J^{b\beta}(p_2) \rangle_{\text{tls}} = &\ \langle t^{\mu \nu}(q) j^{a\alpha}(p_1) j^{b\beta}(p_2) \rangle + \langle t^{\mu \nu}(q) j_{\text{loc}}^{a\alpha}(p_1) j^{b\beta}(p_2) \rangle_g \\
& + \langle t^{\mu \nu}(q) j^{a\alpha}(p_1) j_{\text{loc}}^{b\beta}(p_2) \rangle_g \\
&+ 2 \mathcal{I}^{\mu \nu \rho}(q) \left[ \delta_{[\rho}^{\beta} p_{2 \sigma]} \langle J^{a\alpha}(p_1) J^{b\sigma}(-p_1) \rangle + \delta_{[\rho}^{\alpha} p_{1 \sigma]} \langle J^{b\beta}(p_2) J^{a\sigma}(-p_2) \rangle \right],
\end{aligned}
\label{ali}
\end{equation}
and the trace sector 
\begin{equation}
\langle T^{\mu \nu}(q) J^{a\alpha}(p_1) J^{b\beta}(p_2) \rangle_{\text{tr}} = \frac{1}{3 q^2} \hat{\pi}^{\mu \nu}(q) \left[ \mathcal{A}^{\alpha \beta ab} + \mathcal{B}^{\alpha \beta ab}_g \right].
\label{trc}
\end{equation}
The \(1/q^2\) pole in the equation above is the signature of a dilaton exchanged in the $t$ channel. It is extracted from the longitudinal projector \(\pi^{\mu \nu}\) by defining
\beq
\pi^{\mu\nu}=\frac{1}{q^2 }\hat{\pi}^{\mu \nu}(q) \qquad \hat{\pi}^{\mu \nu}(q) = q^2 g^{\mu \nu} - q^\mu q^\nu .
\eeq
Terms of the form 
\beq
\langle t^{\mu \nu}(q) j_{\text{loc}}^{a\alpha}(p_1) j^{b\beta}(p_2) \rangle_g
\eeq
are new in the tensorial sector decomposition of \eqref{ali}. They are not present in the ordinary $CFT_p$ decomposition of a nonabelian $TJJ $ if the correlator satisfies ordinary Ward identities on the vector currents. As mentioned, in this case they originate from the gluon sector and are allowed by the Slavnov-Taylor identities constraining the gluon currents.\\
The trace contribution includes both the anomaly term  \( \mathcal{A}^{\alpha \beta ab} = \mathcal{A}^{\alpha \beta} \delta^{ab} \), which encodes the conformal anomaly and a second term \( \mathcal{B}_g^{\alpha \beta ab} \), originating from the gluon equations of motion, whose explicit expression is given in \cite{Coriano:2024qbr,Coriano:2025fom}. This term contributes to the trace of the correlator, but it is not part of the anomaly in the strict sense.
The full trace of the correlator takes the form
\beq
g_{\mu \nu} \langle T^{\mu \nu}(q) J^{a\alpha}(p_1) J^{b\beta}(p_2) \rangle_{\text{tr}} = \mathcal{A}^{\alpha \beta ab} + \mathcal{B}_g^{\alpha \beta ab},
\eeq
 where \( \mathcal{A} \) contains a gauge-invariant structure 
 \beq
u^{\alpha \beta} (p_1,p_2)\equiv (p_1\cdot p_2) g^{\alpha\beta} - p_2^\alpha p_1^\beta
\eeq
and is explicitly given by 
\beq
\mathcal{A}^{\alpha \beta ab} = \frac{1}{3} \frac{g_s^2}{16 \pi^2} (11 C_A - 2 n_f) \delta^{ab} u^{\alpha \beta}(p_1, p_2).
\label{onel}
\eeq
This defines the conformal anomaly equation originating both from the gluon and quark sectors, with 
\bea
&&u^{\alpha\beta}(p_1,p_2) = -\frac{1}{4}\int\,d^4x\,\int\,d^4y\ e^{ip_1\cdot x + i p_2\cdot y}\ 
\frac{\delta^2 \{F^a_{\mu\nu}F^{a\mu\nu}(0)\}} {\delta A_{\alpha}(x) A_{\beta}(y)}\vline_{A=0} \,
\label{locvar}\\
\eea
defining the $F^2$ part of the conformal anomaly equation at one-loop level
\begin{equation}
\hat{T}^{\mu}_{\ \mu} = \beta(g) F^{a,\mu\nu} F^a_{\ \mu\nu} + \sum_q m_q \bar{\psi}_q \psi_q \,,
\label{anom1}
\end{equation}
where $\beta(g)$ is the QCD beta function. 
One recognizes in the prefactor of \eqref{onel} the appearance of the leading order QCD $\beta$-function, as expected.
 Therefore the anomaly form factor contains a massless pole whose residue equals $\beta$
\beq
\frac{\beta}{q^2} \delta^{ab} \subset \langle T^{\mu \nu}(q) J^{a\alpha}(p_1) J^{b\beta}(p_2) \rangle. 
\eeq
The contributions from the quark masses have been discussed in \cite{Coriano:2024qbr} and \cite{Coriano:2025fom}. The pole turns into a cut in the presence of mass corrections. A dedicated analysis shows the presence of a nontrivial sum rule \cite{Coriano:2025fom}, satisfied under any kinematical condition by the anomaly form factor 
\beq
\Phi_{\text{an}} \equiv \frac{\mathcal{A}^{\alpha\beta ab}}{q^2}.
\label{ann}
\eeq
In conclusion, the full structure of the \(TJJ\) vertex is more intricate than the one encountered in a purely conformal field theory. It includes both transverse-traceless parts and longitudinal components tied to anomalies and gauge dynamics. The decomposition can be investigated at hadron level using QCD factorization. 

\section{Connecting our parameterization of the $TJJ$ with the hadronic GFF at large $-t$}
We turn to a brief discussion of the connection between the off-shell $TJJ$ vertex derived in the conformal sector and the hadronic matrix element at large momentum transfer $-t$. The focus is on the proton, whose leading Fock state consists of three valence quarks. In the large-$-t$ regime, factorization allows one to relate the matrix element $\langle p', \Lambda' | T^{\mu\nu}_a | p, \Lambda \rangle$ ($a = q, g$) to perturbatively computable partonic amplitudes convoluted with nonperturbative wave functions.\\
The analysis is performed in the Breit frame using a helicity basis, where hadronic states are decomposed into contributions with different orbital angular momentum $l_z$. The dominant component for $A(t)$ arises from helicity-conserving transitions with $l_z = 0$, described by the integrated twist-three light-cone wave function $\Phi_3(x_1, x_2, x_3)$. The GFF $A(t)$ is extracted from the matrix element
\beq
\langle p'_\uparrow | T^{\mu\nu}_a | p_\uparrow \rangle = A_a(t)\, \bar{u}_\uparrow(p') \gamma^{(\mu} P^{\nu)} u_\uparrow(p),
\eeq
with $B(t)$ power suppressed at large $-t$.\\
The hadronic matrix element is written as an overlap of incoming and outgoing three-quark states
\beq
A(t) \bar{u}_\uparrow(p') \gamma^{(\mu} P^{\nu)} u_\uparrow(p) = \int [dx][dy]\, \Phi_3^*(y_i) \Phi_3(x_i)\, \mathcal{T}^{\mu\nu}_{1/2\,1/2}(x_i, y_i),
\eeq
where $\mathcal{T}^{\mu\nu}_{1/2\,1/2}$ represents the partonic matrix element. This structure allows one to compute GFFs from the $TJJ$ vertex through its contribution to the hard scattering.\\
Perturbatively, the leading $O(\alpha_s^2)$ contributions involve insertions of $T_q$ and $T_g$ into tree-level diagrams. As shown in Fig 2, at $O(\alpha_s^3)$, two types of corrections arise: (i) standard radiative corrections involving gluon exchange between valence quarks, and (ii) anomaly-induced corrections from the $TJJ$ vertex that modify gluon propagators.
\begin{figure}[t]
    \centering
    \includegraphics[width=0.6\textwidth]{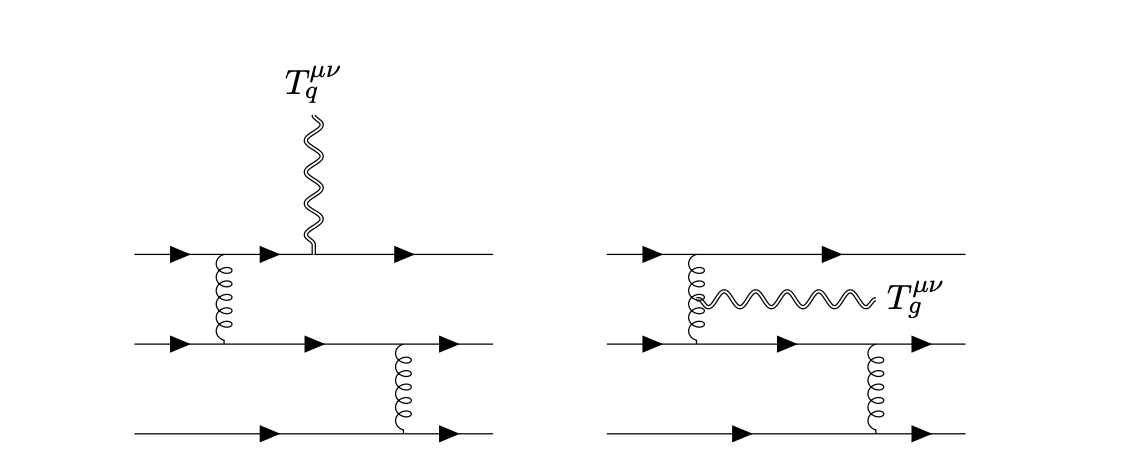}
    \includegraphics[width=0.3\textwidth]{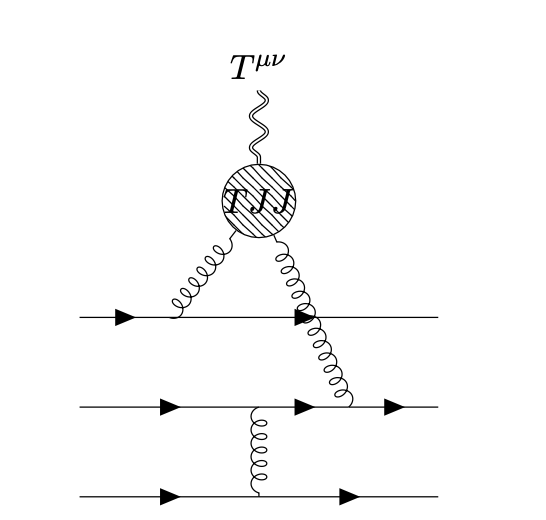}
\caption{Examples of leading $O(\alpha_s^2)$ contributions to the GFF of the proton. Shown are insertions of the graviton/$f \bar{f}$ vertex contained in the quark EMT $T_q$ (left) and the graviton/gg vertex contained in the gluon EMT $T_g$ (center). The complete insertion of the $TJJ$ (right).}    \label{fig:your_label}
\end{figure}
Anomaly effects are introduced through a modified gluon propagator in the hard scattering
\beq
S^{\alpha\beta}(l) \to S^{\alpha\alpha'}(l) \Gamma^{\alpha'\beta'}(q, l, l+q) S^{\beta'\beta}(l+q),
\eeq
with $\Gamma^{\alpha'\beta'}$ encoding the off-shell $TJJ$ structure. These insertions alter the partonic matrix element $\mathcal{T}^{\mu\nu}$. One can use the decomposition discussed above and distill the dilaton contribution 
by a careful analysis of the hadronic transition amplitude. 
The full matrix element, including higher-order and anomaly-induced corrections, is schematically written as
\beq
\langle p'| T^{\mu\nu} | p \rangle = \alpha_s^2 \mathcal{T}^{\mu\nu} + \alpha_s^3 \mathcal{T}_b^{\mu\nu} + \alpha_s^3 \mathcal{T}_{\text{an}}^{\mu\nu},
\eeq
where $\mathcal{T}_b^{\mu\nu}$ represents standard perturbative corrections and $\mathcal{T}_{\text{an}}^{\mu\nu}$ captures the anomaly contribution. These amplitudes are convoluted with hadronic wave functions to yield physical observables.

\begin{figure}[t]
    \centering
    \includegraphics[width=0.4\textwidth]{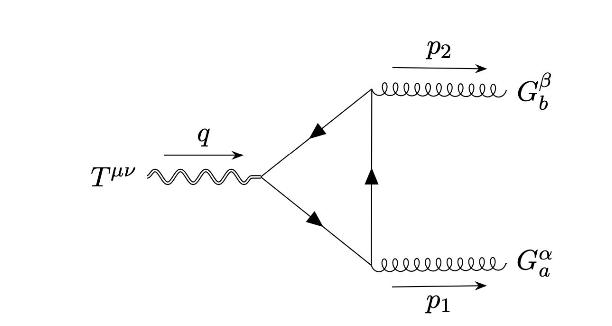}
    \includegraphics[width=0.4\textwidth]{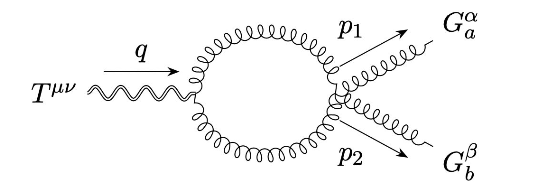}
    \caption{ Examples of typical perturbative contributions with quark and gluons in the $TJJ$. }
    \label{fig:your_label}
\end{figure}

\section{Conclusions}
The analysis we have presented is directly linked to the factorization picture of exclusive processes. This framework naturally led us to consider the perturbative one-loop insertion of the $TJJ$ vertex in the hard scattering. In the hadron case, this insertion contributes at order $\alpha_s^2$, making it subleading compared to the leading $O(\alpha_s)$ corrections from the direct graviton coupling to collinear quarks in the hard scattering. As discussed, the trace part of the $TJJ$ vertex is essentially described by a nonlocal interaction, previously investigated in several anomalous correlators. Its extraction is highly nontrivial and requires the formalism reviewed in this work, based on a combination of general CFT approaches and explicit free-field theory realizations within the $CFT_p$ framework.\\
The structure of the interaction is captured by a longitudinal/transverse/trace decomposition, with the anomaly pole emerging in the trace sector. The hard scattering factorization presented here can be extended to hadronic processes involving the proton and pion, offering a possible phenomenological basis for parameterizing the GFF form factors through invariant amplitudes.\\
Our goal has been to bridge recent developments in the study of anomalies within $CFT_p$ — both in even and odd parity sectors — with the physics of strong interactions. This connection is particularly timely given the renewed focus on anomalies in the context of the Electron-Ion Collider (EIC) program which will play a central role in proton tomography and the determination of the spin and partonic content of hadrons, including anomaly contributions.\\
 Conformal anomalies are inherently linked to the emergence of effective dilaton degrees of freedom in hard scattering processes, with the appearance of an anomaly pole associated with a sum rule, verified perturbatively at one-loop order in the large momentum transfer limit.  
Furthermore, we have shown how the standard \( CFT_p \) approach to three-point functions of conformal correlators can be modified to account for the gauge-fixing sector of gauge theories such as QCD.  
The existence of such sum rules in anomaly form factors is a hallmark of both chiral and conformal anomalies. In the presence of additional scales (such as a quark mass), anomaly poles transform into branch cuts rather than ordinary particle poles, but the corresponding anomaly form factors remain constrained by a sum rule.

\vspace{0.3cm}
\centerline{\bf Acknowledgements}
This work is partially supported by INFN, inziativa specifica {\em QG-sky}, by the the grant PRIN 2022BP52A MUR "The Holographic Universe for all Lambdas" Lecce-Naples, and by the European Union, Next Generation EU, PNRR project "National Centre for HPC, Big Data and Quantum Computing", project code CN00000013.

%\section{ Tensorial Decomposition of the hard scattering}

%\bibliographystyle{jhep}
%\bibliography{TJJdilatonHprime}

\begin{thebibliography}{10}

\bibitem{Polyakov:2018zvc}
M.~V. Polyakov and P.~Schweitzer.
\newblock \,\href{https://doi.org/10.1142/S0217751X18300259}{\emph{Int. J. Mod.
  Phys. A} {\bfseries 33} (2018) 1830025}
  [\href{https://arxiv.org/abs/1805.06596}{{\ttfamily 1805.06596}}].

\bibitem{Teryaev:2016edw}
O.~V. Teryaev.
\newblock \,\href{https://doi.org/10.1007/s11467-016-0573-6}{\emph{Front. Phys.
  (Beijing)} {\bfseries 11} (2016) 111207}.

\bibitem{Burkert:2023wzr}
V.~D. Burkert, L.~Elouadrhiri, F.~X. Girod, C.~Lorc\'e, P.~Schweitzer and P.~E.
  Shanahan.
\newblock \,\href{https://doi.org/10.1103/RevModPhys.95.041002}{\emph{Rev. Mod.
  Phys.} {\bfseries 95} (2023) 041002}
  [\href{https://arxiv.org/abs/2303.08347}{{\ttfamily 2303.08347}}].

\bibitem{Tong:2022zax}
X.-B. Tong, J.-P. Ma and F.~Yuan.
\newblock \,\href{https://doi.org/10.1007/JHEP10(2022)046}{\emph{JHEP}
  {\bfseries 10} (2022) 046}
  [\href{https://arxiv.org/abs/2203.13493}{{\ttfamily 2203.13493}}].

\bibitem{Tanaka:2022wrr}
K.~Tanaka.
\newblock \,\href{https://doi.org/10.7566/JPSCP.37.020405}{\emph{JPS Conf.
  Proc.} {\bfseries 37} (2022) 020405}
  [\href{https://arxiv.org/abs/2209.14367}{{\ttfamily 2209.14367}}].

\bibitem{Hatta:2018sqd}
Y.~Hatta, A.~Rajan and K.~Tanaka.
\newblock \,\href{https://doi.org/10.1007/JHEP12(2018)008}{\emph{JHEP}
  {\bfseries 12} (2018) 008}
  [\href{https://arxiv.org/abs/1810.05116}{{\ttfamily 1810.05116}}].

\bibitem{}
J.~Horejsi.
\newblock \,\href{https://doi.org/10.1007/BF01598423}{\emph{Czech. J. Phys.}
  {\bfseries 42} (1992) 241}.

\bibitem{Ji:1998pc}
X.-D. Ji.
\newblock \,\href{https://doi.org/10.1088/0954-3899/24/7/002}{\emph{J. Phys.}
  {\bfseries G24} (1998) 1181}
  [\href{https://arxiv.org/abs/hep-ph/9807358}{{\ttfamily hep-ph/9807358}}].

\bibitem{Radyushkin:1997ki}
A.~V. Radyushkin.
\newblock \,\href{https://doi.org/10.1103/PhysRevD.56.5524}{\emph{Phys. Rev. D}
  {\bfseries 56} (1997) 5524}
  [\href{https://arxiv.org/abs/hep-ph/9704207}{{\ttfamily hep-ph/9704207}}].

\bibitem{Tarasov:2025mvn}
A.~Tarasov and R.~Venugopalan.
\newblock \,.
\newblock \href{https://arxiv.org/abs/2501.10519}{{\ttfamily 2501.10519}}.

\bibitem{Tarasov:2021yll}
A.~Tarasov and R.~Venugopalan.
\newblock \,\href{https://doi.org/10.1103/PhysRevD.105.014020}{\emph{Phys. Rev.
  D} {\bfseries 105} (2022) 014020}
  [\href{https://arxiv.org/abs/2109.10370}{{\ttfamily 2109.10370}}].

\bibitem{Bhattacharya:2023wvy}
S.~Bhattacharya, Y.~Hatta and W.~Vogelsang.
\newblock \,\href{https://doi.org/10.1103/PhysRevD.108.014029}{\emph{Phys. Rev.
  D} {\bfseries 108} (2023) 014029}
  [\href{https://arxiv.org/abs/2305.09431}{{\ttfamily 2305.09431}}].

\bibitem{Bhattacharya:2022xxw}
S.~Bhattacharya, Y.~Hatta and W.~Vogelsang.
\newblock \,\href{https://doi.org/10.1103/PhysRevD.107.014026}{\emph{Phys. Rev.
  D} {\bfseries 107} (2023) 014026}
  [\href{https://arxiv.org/abs/2210.13419}{{\ttfamily 2210.13419}}].

\bibitem{Castelli:2024eza}
I.~Castelli, A.~Freese, C.~Lorc\'e, A.~Metz, B.~Pasquini and S.~Rodini.
\newblock \,\href{https://doi.org/10.1016/j.physletb.2024.138999}{\emph{Phys.
  Lett. B} {\bfseries 857} (2024) 138999}
  [\href{https://arxiv.org/abs/2408.00554}{{\ttfamily 2408.00554}}].

\bibitem{Ji:1996ek}
X.-D. Ji.
\newblock \,\href{https://doi.org/10.1103/PhysRevLett.78.610}{\emph{Phys. Rev.
  Lett.} {\bfseries 78} (1997) 610}
  [\href{https://arxiv.org/abs/hep-ph/9603249}{{\ttfamily hep-ph/9603249}}].

\bibitem{Radyushkin:1996nd}
A.~V. Radyushkin.
\newblock \,\href{https://doi.org/10.1016/0370-2693(96)00528-X}{\emph{Phys.
  Lett. B} {\bfseries 385} (1996) 333}
  [\href{https://arxiv.org/abs/hep-ph/9604317}{{\ttfamily hep-ph/9604317}}].

\bibitem{Radyushkin:1996ru}
A.~V. Radyushkin.
\newblock \,\href{https://doi.org/10.1016/0370-2693(96)00932-8}{\emph{Phys.
  Lett. B} {\bfseries 380} (1996) 417}
  [\href{https://arxiv.org/abs/hep-ph/9605431}{{\ttfamily hep-ph/9605431}}].

\bibitem{Ji:1996nm}
X.~Ji.
\newblock \,\href{https://doi.org/10.1103/PhysRevD.55.7114}{\emph{Phys. Rev. D}
  {\bfseries 55} (1997) 7114}
  [\href{https://arxiv.org/abs/hep-ph/9609381}{{\ttfamily hep-ph/9609381}}].

\bibitem{Collins:1996fb}
J.~C. Collins and A.~Freund.
\newblock \,\href{https://doi.org/10.1103/PhysRevD.59.074009}{\emph{Phys. Rev.
  D} {\bfseries 59} (1999) 074009}
  [\href{https://arxiv.org/abs/hep-ph/9611369}{{\ttfamily hep-ph/9611369}}].

\bibitem{Vanderhaeghen:1998uc}
M.~Vanderhaeghen, P.~A.~M. Guichon and M.~Guidal.
\newblock \,\href{https://doi.org/10.1103/PhysRevLett.80.5064}{\emph{Phys. Rev.
  Lett.} {\bfseries 80} (1998) 5064}
  [\href{https://arxiv.org/abs/hep-ph/9806305}{{\ttfamily hep-ph/9806305}}].

\bibitem{Kobzarev:1962wt}
I.~Y. Kobzarev and L.~B. Okun.
\newblock \,{\emph{Zh. Eksp. Teor. Fiz.} {\bfseries 43} (1962) 1904}.

\bibitem{Pagels:1966zza}
H.~Pagels.
\newblock \,\href{https://doi.org/10.1103/PhysRev.144.1250}{\emph{Phys. Rev.}
  {\bfseries 144} (1966) 1250}.

\bibitem{Radyushkin:2000uy}
A.~V. Radyushkin.
\newblock \,.
\newblock \href{https://arxiv.org/abs/hep-ph/0101225}{{\ttfamily
  hep-ph/0101225}}.

\bibitem{Goeke:2001tz}
K.~Goeke, M.~V. Polyakov and M.~Vanderhaeghen.
\newblock \,\href{https://doi.org/10.1016/S0146-6410(01)00158-2}{\emph{Prog.
  Part. Nucl. Phys.} {\bfseries 47} (2001) 401}
  [\href{https://arxiv.org/abs/hep-ph/0106012}{{\ttfamily hep-ph/0106012}}].

\bibitem{Diehl:2003ny}
M.~Diehl.
\newblock \,\href{https://doi.org/10.1016/j.physrep.2003.08.002}{\emph{Phys.
  Rept.} {\bfseries 388} (2003) 41}
  [\href{https://arxiv.org/abs/hep-ph/0307382}{{\ttfamily hep-ph/0307382}}].

\bibitem{Belitsky:2005qn}
A.~V. Belitsky and A.~V. Radyushkin.
\newblock \,\href{https://doi.org/10.1016/j.physrep.2005.06.002}{\emph{Phys.
  Rept.} {\bfseries 418} (2005) 1}
  [\href{https://arxiv.org/abs/hep-ph/0504030}{{\ttfamily hep-ph/0504030}}].

\bibitem{Polyakov:1999gs}
M.~V. Polyakov and C.~Weiss.
\newblock \,\href{https://doi.org/10.1103/PhysRevD.60.114017}{\emph{Phys. Rev.
  D} {\bfseries 60} (1999) 114017}
  [\href{https://arxiv.org/abs/hep-ph/9902451}{{\ttfamily hep-ph/9902451}}].

\bibitem{Coriano:2024qbr}
C.~Corian\`o, S.~Lionetti, D.~Melle and R.~Tommasi.
\newblock \,.
\newblock \href{https://arxiv.org/abs/2409.05609}{{\ttfamily 2409.05609}}.

\bibitem{Coriano:2025fom}
C.~Corian\`o, S.~Lionetti, D.~Melle and L.~Torcellini.
\newblock \,.
\newblock \href{https://arxiv.org/abs/2504.01904}{{\ttfamily 2504.01904}}.

\bibitem{Coriano:2025ceu}
C.~Corian\`o, S.~Lionetti and D.~Melle.
\newblock \,.
\newblock \href{https://arxiv.org/abs/2502.03182}{{\ttfamily 2502.03182}}.

\bibitem{Amore:2004ng}
P.~Amore, C.~Coriano and M.~Guzzi.
\newblock \,\href{https://doi.org/10.1088/1126-6708/2005/02/038}{\emph{JHEP}
  {\bfseries 02} (2005) 038}
  [\href{https://arxiv.org/abs/hep-ph/0404121}{{\ttfamily hep-ph/0404121}}].

\bibitem{Coriano:1993mr}
C.~Corian\`o, A.~Radyushkin and G.~F. Sterman.
\newblock \,\href{https://doi.org/10.1016/0550-3213(93)90556-5}{\emph{Nucl.
  Phys. B} {\bfseries 405} (1993) 481}
  [\href{https://arxiv.org/abs/hep-ph/9301274}{{\ttfamily hep-ph/9301274}}].

\bibitem{Coriano:1998ge}
C.~Corian\`o, H.-n. Li and C.~Savkli.
\newblock \,\href{https://doi.org/10.1088/1126-6708/1998/07/008}{\emph{JHEP}
  {\bfseries 07} (1998) 008}
  [\href{https://arxiv.org/abs/hep-ph/9805406}{{\ttfamily hep-ph/9805406}}].

\bibitem{Sterman:1997sx}
G.~F. Sterman and P.~Stoler.
\newblock \,\href{https://doi.org/10.1146/annurev.nucl.47.1.193}{\emph{Ann.
  Rev. Nucl. Part. Sci.} {\bfseries 47} (1997) 193}
  [\href{https://arxiv.org/abs/hep-ph/9708370}{{\ttfamily hep-ph/9708370}}].

\bibitem{Burkert:2018bqq}
V.~D. Burkert, L.~Elouadrhiri and F.~X. Girod.
\newblock \,\href{https://doi.org/10.1038/s41586-018-0060-z}{\emph{Nature}
  {\bfseries 557} (2018) 396}.

\bibitem{Duran:2022xag}
B.~Duran et~al.
\newblock \,\href{https://doi.org/10.1038/s41586-023-05730-4}{\emph{Nature}
  {\bfseries 615} (2023) 813}
  [\href{https://arxiv.org/abs/2207.05212}{{\ttfamily 2207.05212}}].

\bibitem{Belle:2015oin}
{\scshape Belle} collaboration.
\newblock \,\href{https://doi.org/10.1103/PhysRevD.93.032003}{\emph{Phys. Rev.
  D} {\bfseries 93} (2016) 032003}
  [\href{https://arxiv.org/abs/1508.06757}{{\ttfamily 1508.06757}}].

\bibitem{Savinov:2013hda}
{\scshape Belle} collaboration.
\newblock
  \,\href{https://doi.org/10.1016/j.nuclphysbps.2012.12.033}{\emph{Nucl. Phys.
  B Proc. Suppl.} {\bfseries 234} (2013) 287}.

\bibitem{Kumano:2017lhr}
S.~Kumano, Q.-T. Song and O.~V. Teryaev.
\newblock \,\href{https://doi.org/10.1103/PhysRevD.97.014020}{\emph{Phys. Rev.
  D} {\bfseries 97} (2018) 014020}
  [\href{https://arxiv.org/abs/1711.08088}{{\ttfamily 1711.08088}}].

\bibitem{JeffersonLabHallA:2022pnx}
{\scshape Jefferson Lab Hall A} collaboration.
\newblock \,\href{https://doi.org/10.1103/PhysRevLett.128.252002}{\emph{Phys.
  Rev. Lett.} {\bfseries 128} (2022) 252002}
  [\href{https://arxiv.org/abs/2201.03714}{{\ttfamily 2201.03714}}].

\bibitem{CLAS:2022syx}
{\scshape CLAS} collaboration.
\newblock \,\href{https://doi.org/10.1103/PhysRevLett.130.211902}{\emph{Phys.
  Rev. Lett.} {\bfseries 130} (2023) 211902}
  [\href{https://arxiv.org/abs/2211.11274}{{\ttfamily 2211.11274}}].

\bibitem{AbdulKhalek:2021gbh}
R.~Abdul~Khalek et~al.
\newblock \,\href{https://doi.org/10.1016/j.nuclphysa.2022.122447}{\emph{Nucl.
  Phys. A} {\bfseries 1026} (2022) 122447}
  [\href{https://arxiv.org/abs/2103.05419}{{\ttfamily 2103.05419}}].

\bibitem{Li:1992nu}
H.-n. Li and G.~F. Sterman.
\newblock \,\href{https://doi.org/10.1016/0550-3213(92)90643-P}{\emph{Nucl.
  Phys. B} {\bfseries 381} (1992) 129}.

\bibitem{Brodsky:1997de}
S.~J. Brodsky, H.-C. Pauli and S.~S. Pinsky.
\newblock \,\href{https://doi.org/10.1016/S0370-1573(97)00089-6}{\emph{Phys.
  Rept.} {\bfseries 301} (1998) 299}
  [\href{https://arxiv.org/abs/hep-ph/9705477}{{\ttfamily hep-ph/9705477}}].

\bibitem{Matveev:1973ra}
V.~A. Matveev, R.~M. Muradian and A.~N. Tavkhelidze.
\newblock \,\href{https://doi.org/10.1007/BF02728133}{\emph{Lett. Nuovo Cim.}
  {\bfseries 7} (1973) 719}.

\bibitem{Brodsky:1973kr}
S.~J. Brodsky and G.~R. Farrar.
\newblock \,\href{https://doi.org/10.1103/PhysRevLett.31.1153}{\emph{Phys. Rev.
  Lett.} {\bfseries 31} (1973) 1153}.

\bibitem{Ji:2003fw}
X.-d. Ji, J.-P. Ma and F.~Yuan.
\newblock \,\href{https://doi.org/10.1103/PhysRevLett.90.241601}{\emph{Phys.
  Rev. Lett.} {\bfseries 90} (2003) 241601}
  [\href{https://arxiv.org/abs/hep-ph/0301141}{{\ttfamily hep-ph/0301141}}].

\bibitem{Ji:2002xn}
X.-d. Ji, J.-P. Ma and F.~Yuan.
\newblock \,\href{https://doi.org/10.1016/S0550-3213(03)00010-5}{\emph{Nucl.
  Phys. B} {\bfseries 652} (2003) 383}
  [\href{https://arxiv.org/abs/hep-ph/0210430}{{\ttfamily hep-ph/0210430}}].

\bibitem{Ji:2003yj}
X.-d. Ji, J.-P. Ma and F.~Yuan.
\newblock \,\href{https://doi.org/10.1140/epjc/s2003-01563-y}{\emph{Eur. Phys.
  J. C} {\bfseries 33} (2004) 75}
  [\href{https://arxiv.org/abs/hep-ph/0304107}{{\ttfamily hep-ph/0304107}}].

\bibitem{Lepage:1980fj}
G.~P. Lepage and S.~J. Brodsky.
\newblock \,\href{https://doi.org/10.1103/PhysRevD.22.2157}{\emph{Phys. Rev. D}
  {\bfseries 22} (1980) 2157}.

\bibitem{Efremov:1979qk}
A.~V. Efremov and A.~V. Radyushkin.
\newblock \,\href{https://doi.org/10.1016/0370-2693(80)90869-2}{\emph{Phys.
  Lett. B} {\bfseries 94} (1980) 245}.

\bibitem{Coriano:2020ees}
C.~Corian\`o and M.~M. Maglio.
\newblock \,\href{https://doi.org/10.1016/j.physrep.2021.11.005}{\emph{Phys.
  Rept.} {\bfseries 952} (2022) 2198}
  [\href{https://arxiv.org/abs/2005.06873}{{\ttfamily 2005.06873}}].

\bibitem{Bzowski:2013sza}
A.~Bzowski, P.~McFadden and K.~Skenderis.
\newblock \,\href{https://doi.org/10.1007/JHEP03(2014)111}{\emph{JHEP}
  {\bfseries 03} (2014) 111} [\href{https://arxiv.org/abs/1304.7760}{{\ttfamily
  1304.7760}}].

\bibitem{Bzowski:2018fql}
A.~Bzowski, P.~McFadden and K.~Skenderis.
\newblock \,\href{https://doi.org/10.1007/JHEP11(2018)159}{\emph{JHEP}
  {\bfseries 11} (2018) 159}
  [\href{https://arxiv.org/abs/1805.12100}{{\ttfamily 1805.12100}}].

\end{thebibliography}

\providecommand{\href}[2]{#2}\begingroup\raggedright\endgroup

\end{document}